\documentclass[manuscript]{aastex62}

\usepackage[utf8]{inputenc}
\usepackage[T1]{fontenc}

\graphicspath{{./}{figures/}}

\received{September 6, 2019}
\accepted{October 10, 2019}

\shorttitle{Parker Solar Probe Electrons}
\shortauthors{Halekas et al.}

\begin{document}

\title{Electrons in the Young Solar Wind: First Results from the Parker Solar Probe}

\correspondingauthor{Jasper S. Halekas}
\email{jasper-halekas@uiowa.edu}

\author[0000-0001-5258-6128]{J.~S. Halekas}
\affil{Department of Physics and Astronomy, 
University of Iowa, 
Iowa City, IA 52242, USA}

\author[0000-0002-7287-5098]{P. Whittlesey}
\affil{Space Sciences Laboratory, University of California, Berkeley, CA 94720, USA}

\author[0000-0001-5030-6030]{D.~E. Larson}
\affil{Space Sciences Laboratory, University of California, Berkeley, CA 94720, USA}

\author[0000-0002-6405-7415]{D. McGinnis}
\affil{Department of Physics and Astronomy, 
University of Iowa, 
Iowa City, IA 52242, USA}

\author[0000-0001-6172-5062]{M. Maksimovic}
\affil{LESIA, Observatoire de Paris, Universite PSL, CNRS, Sorbonne Universite, Universite de Paris, 5 place Jules Janssen, 92195 Meudon, France}

\author[0000-0001-6235-5382]{M. Berthomier}
\affil{Laboratoire de Physique des Plasmas, CNRS, Sorbonne Universite, Ecole Polytechnique, Observatoire de Paris, Universite Paris-Saclay, Paris, 75005, France}

\author[0000-0002-7077-930X]{J.~C. Kasper}
\affil{Climate and Space Sciences and Engineering, University of Michigan, Ann Arbor, MI 48109, USA}
\affil{Smithsonian Astrophysical Observatory, Cambridge, MA 02138, USA}

\author[0000-0002-3520-4041]{A.~W. Case}
\affil{Smithsonian Astrophysical Observatory, Cambridge, MA 02138, USA}

\author[0000-0001-6095-2490]{K.~E. Korreck}
\affil{Smithsonian Astrophysical Observatory, Cambridge, MA 02138, USA}

\author[0000-0002-7728-0085]{M.~L. Stevens}
\affil{Smithsonian Astrophysical Observatory, Cambridge, MA 02138, USA}

\author[0000-0001-6038-1923]{K.~G. Klein}
\affil{Department of Planetary Sciences, University of Arizona, Tucson, AZ 85721 USA}

\author[0000-0002-1989-3596]{S.~D. Bale}
\affil{Space Sciences Laboratory, University of California, Berkeley, CA 94720, USA}
\affil{Physics Department, University of California, Berkeley, CA 94720, USA}

\author[0000-0003-3112-4201]{R.~J. MacDowall}
\affil{NASA/Goddard Space Flight Center, Greenbelt, MD 20771, USA}

\author[0000-0002-1573-7457]{M.~P. Pulupa}
\affil{Space Sciences Laboratory, University of California, Berkeley, CA 94720, USA}

\author[0000-0003-1191-1558]{D.~M. Malaspina}
\affil{Astrophysical and Planetary Sciences Department, University of Colorado, Boulder, CO 80309, USA}
\affil{Laboratory for Atmospheric and Space Physics, University of Colorado, Boulder, CO 80303, USA}

\author[0000-0003-0420-3633]{K. Goetz}
\affiliation{School of Physics and Astronomy, University of Minnesota, Minneapolis, MN 55455, USA}

\author[0000-0002-6938-0166]{P.~R. Harvey}
\affil{Space Sciences Laboratory, University of California, Berkeley, CA 94720, USA}

\begin{abstract}

The Solar Wind Electrons Alphas and Protons experiment on the Parker Solar Probe (PSP) mission measures the three-dimensional electron velocity distribution function. We derive the parameters of the core, halo, and strahl populations utilizing a combination of fitting to model distributions and numerical integration for  $\sim 100,000$ electron distributions measured near the Sun on the first two PSP orbits, which reached heliocentric distances as small as $\sim 0.17$ AU. As expected, the electron core density and temperature increase with decreasing heliocentric distance, while the ratio of electron thermal pressure to magnetic pressure ($\beta_e$) decreases. These quantities have radial scaling consistent with previous observations farther from the Sun, with superposed variations associated with different solar wind streams. The density in the strahl also increases; however, the density of the halo plateaus and even decreases at perihelion, leading to a large strahl/halo ratio near the Sun. As at greater heliocentric distances, the core has a sunward drift relative to the proton frame, which balances the current carried by the strahl, satisfying the zero-current condition necessary to maintain quasi-neutrality. Many characteristics of the electron distributions near perihelion have trends with solar wind flow speed, $\beta_e$, and/or collisional age. Near the Sun, some trends not clearly seen at 1 AU become apparent, including anti-correlations between wind speed and both electron temperature and heat flux. These trends help us understand the mechanisms that shape the solar wind electron distributions at an early stage of their evolution.  

\end{abstract}


\section{Introduction} 
\label{sec:intro}

The solar wind that flows out from the Sun contains an  admixture of ions and electrons, with charge neutrality maintained on all macroscopic scales within the wind. Intuitively, the simplest such solution would consist of a thermal distribution of electrons with the same bulk velocity as the solar wind ions. However, observations instead reveal that solar wind electrons have complex distribution functions, with not only a thermal core population, but also a suprathermal halo population, and often an additional suprathermal magnetic field-aligned population known as the strahl \citep{feldman_solar_1975, pilipp_characteristics_1987, rosenbauer_survey_1977, maksimovic_radial_2005, stverak_radial_2009}. These populations satisfy not only local quasi-neutrality ($\sum Z_i n_i = n_e$), but also the zero current condition ($\sum Z_i n_i v_i = n_e v_e$) required to ensure quasi-neutrality globally, with these conditions typically achieved by opposite drifts of the core and suprathermal populations with respect to the ion bulk flow \citep{feldman_solar_1975, scime_regulation_1994, stverak_radial_2009}. 

In order to maintain quasi-neutrality in the presence of mass-dependent gravitational forces, electric fields must exist \citep{pannekoek_ionization_1922}. Given the large thermal velocity of the light electrons, a considerable electric potential drop should exist between the solar corona and 1 AU \citep{lemaire_kinetic_1971}. A class of "exospheric" models \citep{lemaire_kinetic_1973} posits that this electric field accelerates the solar wind ions from the corona, with the pervasive non-thermal nature of the electron distribution increasing the efficiency of this acceleration \citep{pierrard_lorentzian_1996, maksimovic_kinetic_1997, scudder_causes_1992}. Regardless of whether these models prove correct, a significant sunward electric field must exist in the solar wind, and it should play a role in modifying the distributions of the electrons as they escape the corona. This electric field may self-consistently generate the non-thermal features of the electron distribution through a runaway process \citep{Scudder:SERM}.

As electrons travel outward from the Sun, they experience not only an electric force, but also a magnetic force from the interplanetary magnetic field (IMF). Given the conservation of the first adiabatic invariant $\mu = \frac{1}{2} m v_{\perp}^2 / B$, electrons would spiral outward along the magnetic field with a decreasing pitch angle between the electron velocity and the magnetic field, with their velocities becoming increasingly magnetic field-aligned with distance from the Sun.  

However, in addition to forces due to quasi-static fields, wave-particle interactions and Coulomb collisions can affect the solar wind electrons, and both can break the conservation of $\mu$. The non-thermal and anisotropic nature of the electron distribution can drive a wide variety of instabilities \citep{gary_heat_1975, gary_electron_1999, tong_effects_2015, vasko_whistler_2019}, which may in turn modify the distribution function \citep{vocks_electron_2005}. At the same time, Coulomb collisions must occur, and these also shape the electron distribution \citep{scudder_theory_1979, scudder_theory_1979-1, salem_electron_2003}. Electrons in the thermal core can experience many collisions between the corona and 1 AU, while those in the suprathermal strahl and halo experience fewer collisions, thanks to the steep velocity dependence of the Coulomb cross section. 

The electron halo and strahl evolve with increasing radial distance from the Sun, with the fraction of the distribution in the halo increasing, and the fraction of the distribution in the strahl decreasing, suggesting that some process(es) may transform the strahl into the halo \citep{maksimovic_radial_2005, stverak_radial_2009}. At the same time, the strahl angular width increases with radial distance \citep{hammond_variation_1996, graham_evolution_2017, bercic_scattering_2019}, rather than decreasing as expected for adiabatic behavior. These characteristics likely indicate the result of wave-particle interactions and/or Coulomb collisions acting on the electron distribution. 

The Helios mission returned measurements of solar wind electrons from as close to the Sun as $\sim0.3$ AU. At this distance, measured electron distributions already displayed pervasive non-thermal characteristics and anisotropy with respect to the magnetic field \citep{rosenbauer_survey_1977}. The origins of these features remain poorly understood. 

Closer to the Sun, the electrons may approach a  state closer to the initial distribution function in the outer corona, less modified by collisions or instabilities, allowing us to improve our understanding of the mechanisms that form the electron distributions we observe at greater heliocentric distances. In this paper we describe electron observations made by the Solar Wind Electrons Alphas and Protons (SWEAP) experiment \citep{kasper_solar_2016} on the first two orbits of the Parker Solar Probe (PSP) mission \citep{fox_solar_2016}, reaching heliocentric distances as small as $\sim0.17$ AU. 

\section{Electron Distributions} \label{sec:data}

The SWEAP suite contains two Solar Probe Analyzer (SPAN) electron sensors on the ram (Ahead) and anti-ram (Behind) faces of the spacecraft (SPAN-A-E and SPAN-B-E) , which together measure the majority of the three-dimensional electron velocity distribution function \citep{Whittlesey:SPANe}. Each sensor utilizes electrostatic deflectors, toroidal electrostatic analyzers, and microchannel plate detectors above a segmented anode to measure electrons over a $\sim 240^{\circ} \times 120^{\circ}$ field of view (FOV). For the first two PSP orbits, the SPAN high voltage sweep was set to measure electrons over an energy range of $\sim 2-2000$ eV, sufficient to cover the core, halo, and strahl. The measurements utilized in this manuscript have angular resolution of at least $ 24^{\circ} \times \sim 20^{\circ}$, with higher resolution of $ 6^{\circ} \times \sim 20^{\circ}$ for some sunward look directions (with the first angular dimension fixed by the anode size, and the second varying over the deflector sweep). Combined, the two sensor FOVs cover the majority of the sky; however, some gaps remain, necessitating either interpolation or fitting to compute accurate moments. In this work, we elect to use a combination of fitting and numerical integration to determine the electron properties, as described in Appendix \ref{sec:appen}.  

Figure \ref{fig:f1} shows a sample electron distribution measured by SPAN-A-E and SPAN-B-E near perihelion on PSP's first orbit, transformed to the solar wind proton frame estimated from moments computed from Solar Probe Cup (SPC) measurements \citep{Case:SPC}. The observed distribution contains four populations with different energies and angular extents. At larger energies, we see a relatively isotropic core and halo with a break point at velocities of $\sim 10^4$ km/s, and an anti-field-aligned strahl covering a narrow angular range, with velocities $\gtrsim 5000$ km/s. At the highest energies, the strahl has phase space densities orders of magnitude higher than those in other directions, as seen in the top panels of Fig. \ref{fig:f1}. At smaller energies, on the other hand, relative variations in flux do not exceed a factor of two, as seen in the middle panels of Fig. \ref{fig:f1}. We also note the presence of partial obstructions by portions of the spacecraft, apparent at lower energies, in four pixels with near-sunward look directions.   

\begin{figure}
\epsscale{0.85}
\plotone{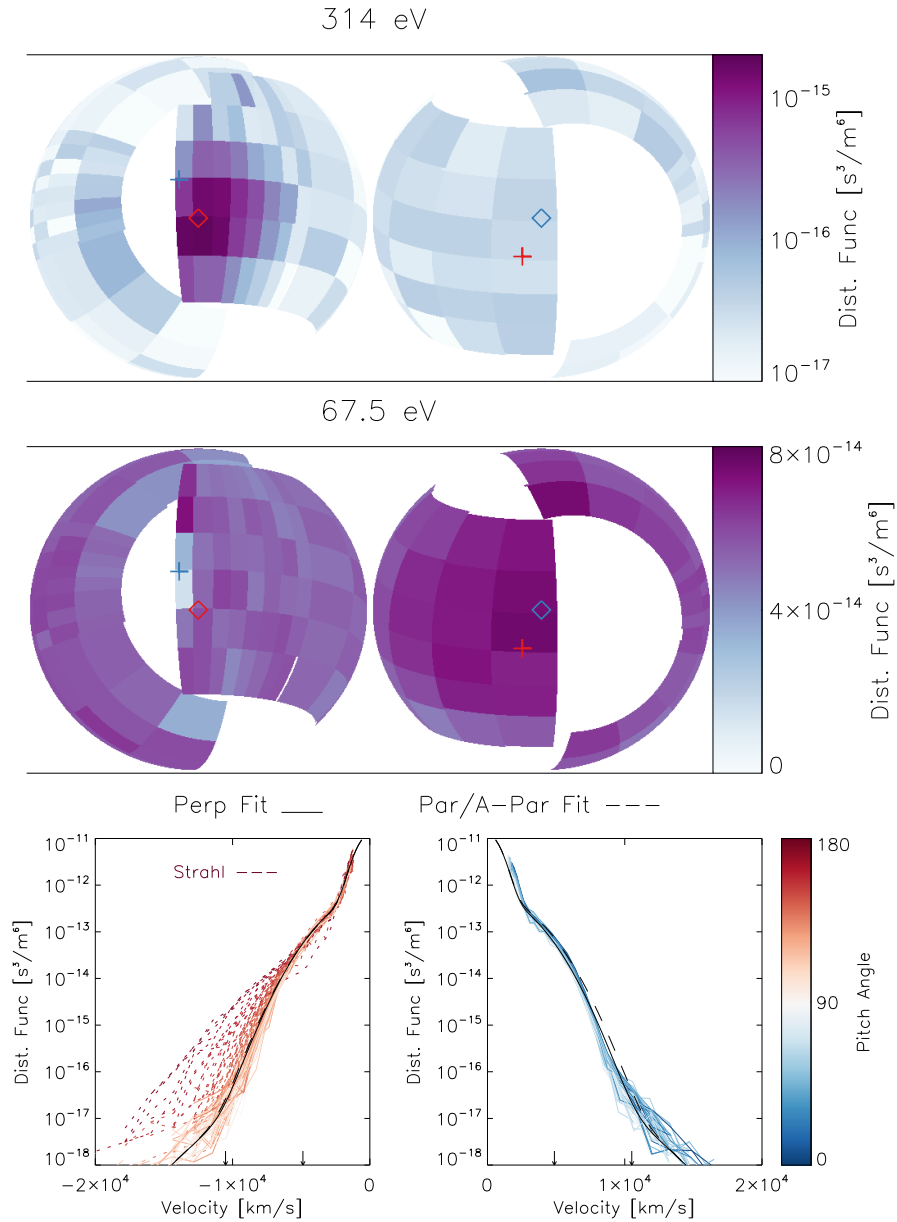}
\caption{Electron distribution function from 12:56:29 UT on 06 Nov 2018, in the proton frame. The top four panels show orthographic field-of-view (FOV) projections. Diamonds and pluses indicate velocities parallel (blue) and anti-parallel (red) to the magnetic field and the proton velocity. The bottom panels show measurements from each look direction. Arrows indicate the velocities of the two FOV plots. Black solid and dashed lines show cuts through a model secondary+core+halo distribution fit to the non-strahl pitch angles.  \label{fig:f1}}
\end{figure}

A fourth population at the lowest measured energies has characteristics consistent with secondary electrons produced from instrument and/or spacecraft surfaces, with a higher relative flux at times with higher core electron temperatures, as often seen in regions of the terrestrial magnetosphere with high electron temperature \citep{mcfadden_themis_2009}. Unlike photoelectrons produced from spacecraft and/or instrument surfaces, which also commonly affect electron measurements \citep{scime_effects_1994}, we find no clear delineation in energy between this population and the core. This provides an additional motivation to use a fitting procedure rather than direct integration to characterize the electron core. 

For the distribution shown in Fig. \ref{fig:f1}, utilizing the analysis procedure described in Appendix \ref{sec:appen}, we find best-fit values of core density $n_c = 329 cm^{-3}$, core temperatures $kT_{||} =30.9$ eV and $kT_{\perp} = 29.3$ eV, core drift speed $v_c = 152$ km/s, halo density $n_h =0.58 cm^{-3}$, and halo temperature $kT_h = 124$ eV. The core density values agree with measurements from SPC \citep{Case:SPC} and from quasi-thermal noise measurements \citep{Moncuquet:QTN}. We numerically compute the strahl moments as described in Appendix \ref{sec:appen}, finding a partial density $n_s = 4.0 cm^{-3}$, average parallel velocity $<v_s> = -6910$ km/s, average energy $<E_s> = 173$ eV, and heat flux (energy flux in the proton frame) $Q_{s}=-8.7 \times 10^{-4} W/m^2$, where the negative signs of the odd moments indicate anti-alignment to the nearly sunward magnetic field measured by FIELDS \citep{bale_fields_2016}. 

Despite the high amplitude of the strahl, it has very limited angular extent, and the suprathermal density fraction $(n_s + n_h)/n_c \sim 0.015$ is smaller than typical values of $\sim 0.04-0.08$ \citep{feldman_solar_1975, stverak_radial_2009} at greater heliocentric distances. Similarly, we find a suprathermal energy density fraction $(n_s <E_s> + \frac{3}{2}n_h k T_h)/(\frac{3}{2} n_c k T_c) \sim 0.05$ significantly lower than reported values of $\sim 0.2-0.4$ at 1 AU \citep{feldman_solar_1975}. While underestimates of the strahl density due to FOV gaps play a role for some measurements, they cannot explain the magnitude of the difference. At this time, we cannot determine whether the reduced suprathermal fraction is characteristic of the near-Sun environment, or particular to the limited range of heliographic latitude and longitude sampled near perihelion on the first two PSP orbits.  

Near perihelion, we commonly find a deficit in the measured electron distribution with respect to the best-fit model distribution in the anti-strahl direction, as seen in Fig. \ref{fig:f1}. This additional source of skew in the suprathermal electron distribution may result from the same mechanisms that create the strahl, or it may arise from the initial conditions at the corona and/or the effects of the electric fields discussed in Section \ref{sec:intro}. For an outward-directed coronal source of electrons, the net electric potential drop along the magnetic field (on the order of hundreds of volts, comparable to the energy of the sunward deficit) leads to a natural distinction between trapped and escaping electrons, which could create a truncation of the sunward-directed portion of the distribution \citep{lemaire_kinetic_1971, pierrard_electron_1999}. The presence of a sunward halo population above the velocity range of the observed deficit may argue against this concept; however, the halo may form by a different set of processes, such as wave-particle interactions.

For the distribution of Fig. \ref{fig:f1}, we find an estimated core drift current $e n_c v_c = 8.0 \mu A/m^2$, which somewhat exceeds the magnitude of the computed strahl current $e n_s <v_s> = -4.5 \mu A/m^2$. However, when we numerically integrate the deficit in the anti-strahl direction, as described in Appendix \ref{sec:appen}, we find a density decrement of $-2.5 cm^{-3}$, with a corresponding velocity moment of $3110$ km/s. After adding this to the best-fit core values, we calculate an effective core drift velocity $v_c (*) = 107$ km/s and current $e n_c v_c(*) = 5.6 \mu A/m^2$, closer to that required to balance the strahl current. Much of the remaining imbalance likely results from unmeasured strahl current due to FOV gaps, given the angular location of the strahl near the spacecraft heat shield at this time. However, measurement errors and/or misfits may also play a role.

\section{Electron Properties Observed on Orbits 1 and 2} \label{sec:obs}

We repeat the analysis of Section \ref{sec:data} for 125,401 individual distribution functions measured by the SPAN sensors between 2 Oct 2018 and 11 Apr 2019. Of these, we find 99,049 distributions for which we can retrieve the core properties successfully, with the excluded fits having an unacceptable chi-square value and/or returning un-physical parameters. We find 59,213 distributions for which we can retrieve the halo/strahl parameters successfully, with the majority of the additional excluded cases due to insufficient counts in the suprathermal portion of the distribution, both at larger distances from the Sun (lower densities) and near perihelion on Orbit 2 (moderate densities, but sensor mechanical attenuators closed, reducing the instrumental sensitivity).   

\begin{figure}
\plotone{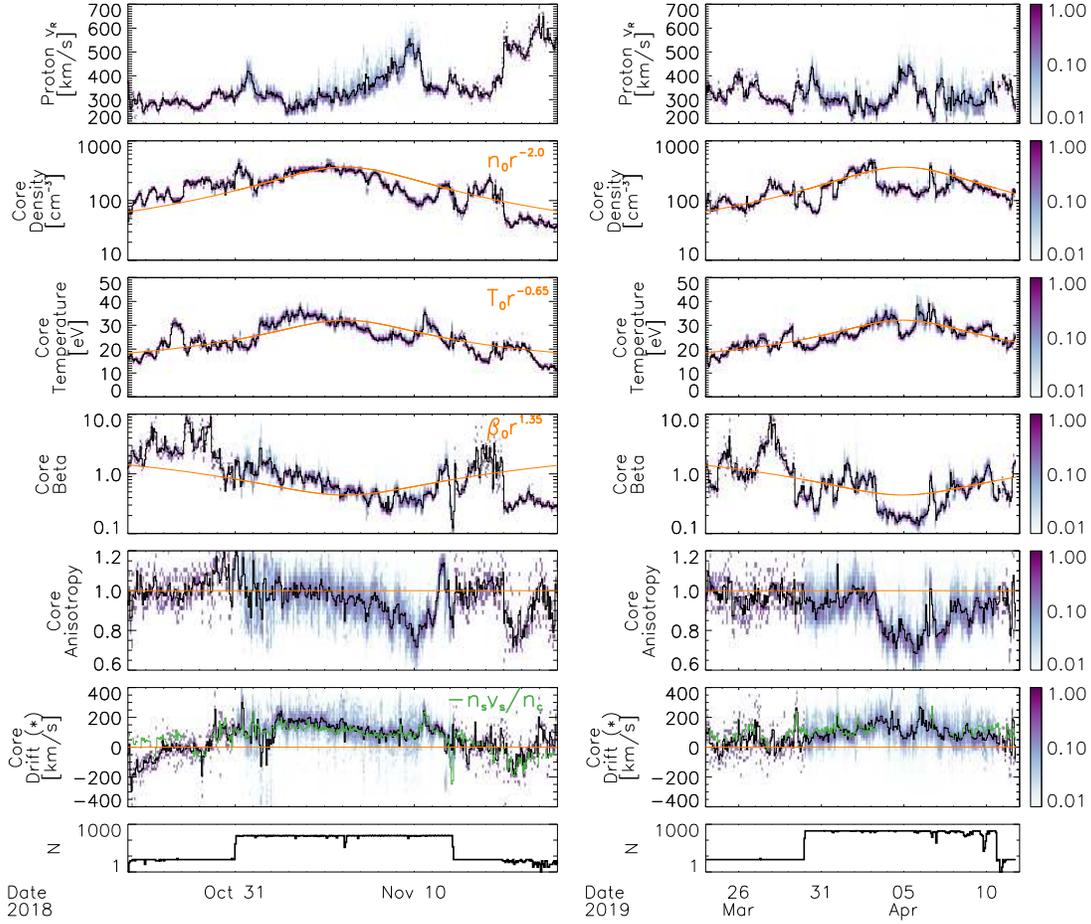}
\caption{Distribution function parameters for the electron core, for a portion of orbits 1 and 2 surrounding perihelion, within 0.4 AU. The first six panels show normalized frequency histograms in color for each $\sim 1.5$ hour interval for the radial proton speed $v_R$ and the electron core density $n_c$, temperature $kT_c = (2 kT_{\perp}+kT_{||})/3$, thermal to magnetic pressure ratio $\beta_c$, temperature anisotropy $T_{\perp}/T_{||}$, and corrected drift speed $v_c (*)$ parallel or anti-parallel to the magnetic field. Black lines on these panels indicate the average values at each time. The bottom panel shows the number of individual measurements per histogram. Orange lines indicate simple radial scaling predictions. The green line on the sixth panel shows the core drift needed to balance the average measured strahl current, in the proton frame. \label{fig:f2}}
\end{figure}

\begin{figure}
\plotone{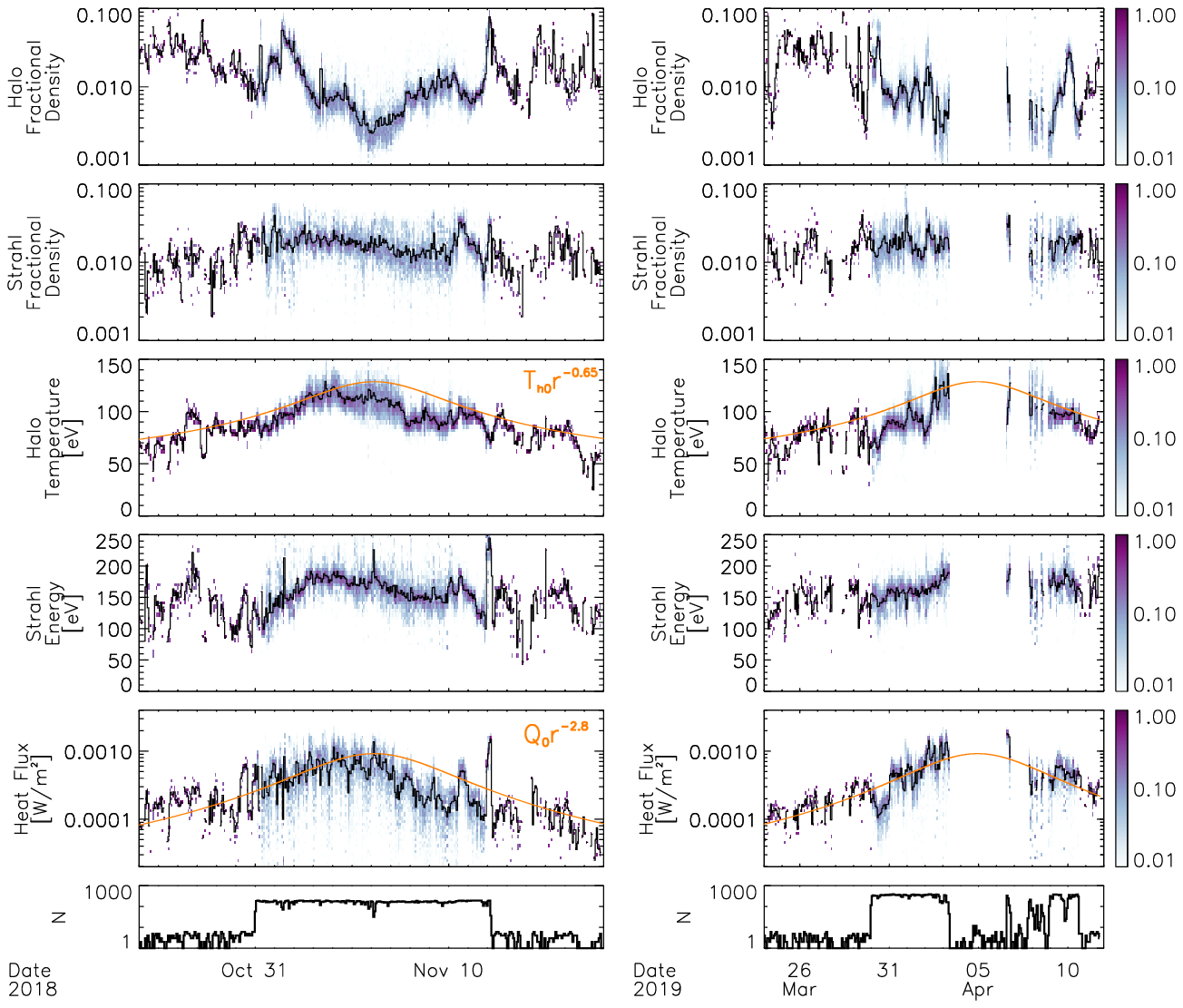}
\caption{Distribution function parameters for the electron halo and strahl, in the same format as Fig. \ref{fig:f2}, for the same time periods. The first five panels show halo and strahl fractional density $n_h/n_c$ and $n_s/n_c$, halo temperature $kT_h$, strahl average energy $<E_s>$, and total electron heat flux $Q$. \label{fig:f3}}
\end{figure}

Figures \ref{fig:f2} and \ref{fig:f3} summarize the results of this analysis, for distributions measured within 0.4 AU on the first two PSP orbits. As discussed by \citep{Kasper2019:prep}, PSP nearly co-rotates with the Sun near perihelion, and so its orbit traverses only a small range of heliographic latitude and longitude. Therefore, PSP samples only a limited subset of solar wind sources, so we should not necessarily consider the results representative of the average radial scaling of the observed parameters. Nonetheless, the observed density roughly follows the expected $r^{-2}$ scaling of a spherically expanding wind, the core and halo temperatures scale consistently with observations at greater heliocentric distances \citep{maksimovic_radial_2005}, and $\beta_c$ has trends consistent with the scaling of these quantities and the $\sim r^{-2}$ scaling of the magnetic field strength appropriate for a nearly radial field. 

However, superposed on the radial scaling, we find clear variations in the electron properties, which correlate with different solar wind streams, as seen in Fig. \ref{fig:f2}. During Orbit 1, we observe a clear evolution across the perihelion encounter. As the radial speed increases from $\sim 300$ to $\sim 500$ km/s, the core density, core and halo temperatures, and $\beta_c$ all decrease in comparison to radial scaling. At the same time, the core becomes more anisotropic, the strahl fractional density decreases, the core drift speed decreases, and the heat flux decreases. These trends may result from connection to different portions of a coronal hole \citep{Bale:prep}. We find generally similar trends associated with a $\sim 350-400$ km/s solar wind stream encountered near perihelion on Orbit 2. 

At most times, core drift speeds approximately balance the strahl electron flux in the proton frame and satisfy the zero-current condition, as seen in Fig. \ref{fig:f2}. For the majority of both orbits, and during both encounters, we find positive (parallel to the magnetic field) core drift values, consistent with the prevailing sunward magnetic field \citep{Bale:prep}. To make this comparison, we corrected core drift speeds for the anti-strahl deficit as described in Section \ref{sec:data}, and excluded times when the predicted strahl direction fell in the largest FOV gap. The remaining discrepancies likely result from small misfits of the core, which can result from the contamination by secondary electrons described in Section \ref{sec:data}, though our multi-component fitting technique helps mitigate these effects. In addition, spacecraft potential may affect the results, though preliminary estimated values were small (< $\pm$ 5 V) during Orbits 1 and 2. Finally, the effects of alpha particles could play a role in offsetting the true plasma frame from the proton frame.

Consistent with expectations based on previous investigations, we find very low halo fractional densities near perihelion, much smaller than at larger heliocentric distances \citep{mccomas_solar_1992}, and considerably smaller even than those previously reported at 0.3 AU \citep{maksimovic_radial_2005, stverak_radial_2009}. The strahl fractional density appears roughly constant, on the order of $\sim 0.01-0.03$, somewhat smaller than but not obviously inconsistent with previous work, particularly given our limited sampling of solar wind sources. 

We compute the total electron heat flux in the proton frame by combining the sunward heat flux estimated from the corrected core fit parameters as in \cite{feldman_solar_1975} with the anti-sunward heat flux from the strahl moments. We find trends consistent with previous measurements \citep{pilipp_large-scale_1990,scime_regulation_1994}, with a similar radial exponent to that previously found for the slow solar wind \citep{stverak_electron_2015}, though with somewhat lower magnitudes, consistent with the relatively small strahl fractional densities we observe on these first two orbits. The heat flux radial scaling, steeper than the $r^{-2}$ expected for expansion along a near-radial field, is consistent with the temperature scaling, less steep than the $r^{-4/3}$ expected from a purely adiabatic radial expansion \citep{maksimovic_radial_2005, stverak_electron_2015}. This degradation of the electron heat flux likely results from the same mechanism(s) that modify the strahl and halo with increasing heliocentric distance. 

\section{Trends in Electron Properties Within 0.25 AU} \label{sec:props}

\begin{figure}
\plotone{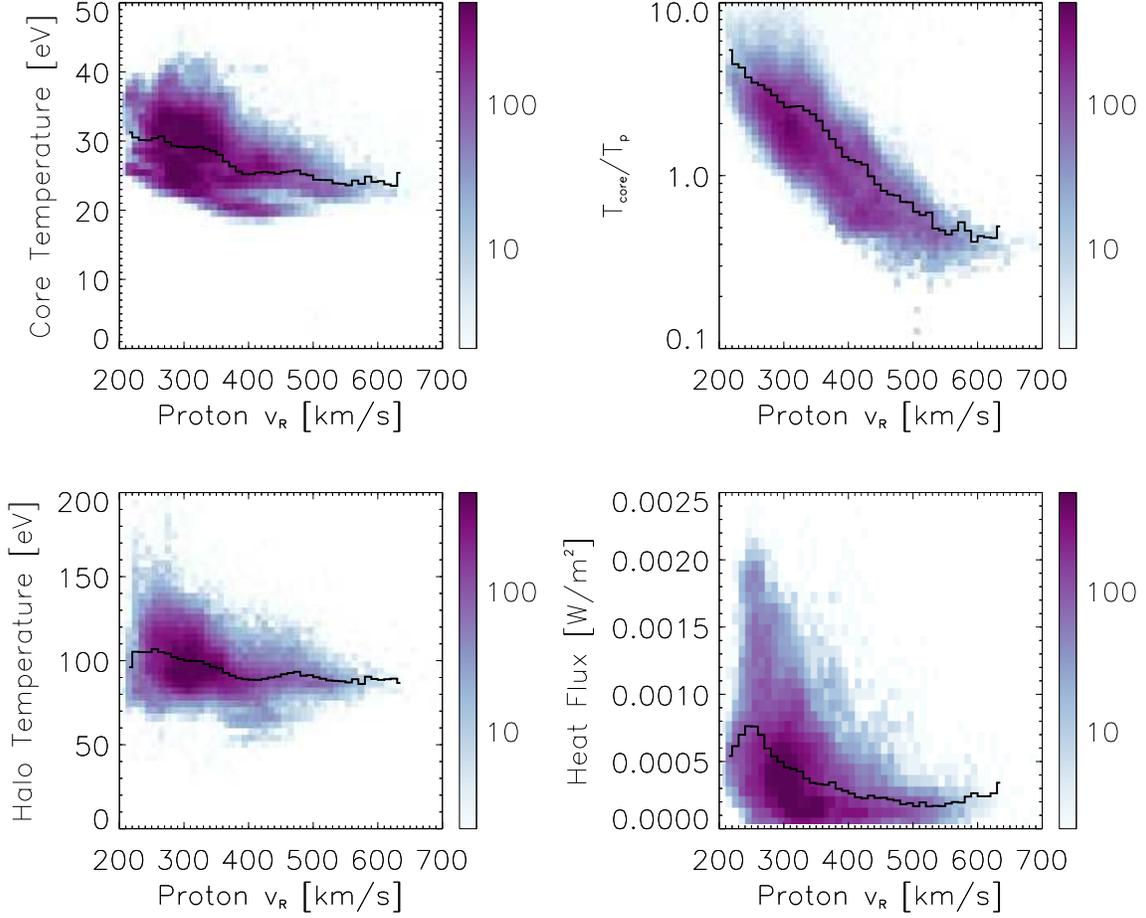}
\caption{Frequency distributions of selected distribution function parameters vs. radial flow speed for all measurements within 0.25 AU. Colors indicate the number of individual measurements in each bin (normalized by the bin size for logarithmic bins to properly represent the data density). Black lines show the average values of the parameters for each flow speed bin.  \label{fig:f4}}
\end{figure}

We investigate the trends in electron properties as a function of selected solar wind parameters. To explore a region not previously measured, and to focus on time periods with the highest measurement cadence and data density, we focus our remaining analysis on heliocentric distances less than 0.25 AU. We first consider trends with respect to the solar wind radial flow speed $v_R$ measured by SPC, with selected results shown in Figure \ref{fig:f4}. 

At 1 AU, the flow speed of the wind organizes many properties of the solar wind plasma, with high speed wind having typically lower density, higher proton temperature, and higher Alfv\'{e}nicity \citep{hundhausen_solar_1970, belcher_large-amplitude_1971}. The high speed wind also has lower minor ion charge state ratios, indicating a lower coronal electron temperature at the freeze-in point \citep{geiss_southern_1995, gloeckler_implications_2003}, suggesting an origin in cooler coronal holes. However, by 1 AU, any correlation between solar wind speed and in situ electron temperature has typically vanished, possibly destroyed by compressive mechanisms such as stream-stream interactions. 

As seen in Fig. \ref{fig:f4}, we find a clear anti-correlation between solar wind speed and both core and halo temperatures near perihelion, suggesting that within 0.25 AU the in situ electrons still retain a memory of their origin in the corona. Similar results based on a reanalysis of Helios data are also reported in this volume \citep{Maxsimovic:Helios_comp}. 

%% We observe an anti-correlation between core temperature anisotropy and flow speed, as seen in Fig. \ref{fig:f4}. Given that high speed solar wind typically has a lower collisional age, we expect a less isotropic distribution in the fast wind. This trend also appears clearly at greater heliocentric distances \citep{phillips_anisotropic_1989, salem_electron_2003}.

Since the proton temperature still correlates with the flow speed as at 1 AU, the ratio between electron and proton temperatures has a very strong anti-correlation with flow speed near perihelion. The slowest wind has electron temperatures significantly larger than proton temperatures, while the higher speed wind has proton temperatures larger than electron temperatures. These trends have implications for the growth of instabilities \citep{Klein:prep} as well as for the heating of the solar wind near the Sun. 

We also find an anti-correlation between electron heat flux and flow speed, much more apparent than at greater heliocentric distances, where at best weak trends appear for even carefully selected observations \citep{salem_electron_2003}. In fact, at 1 AU the fast wind is typically thought to contain electron distributions with more significant non-thermal features \citep{feldman_characteristic_1978}, though the average difference in the fractional abundance of suprathermals is not large \citep{stverak_radial_2009}. The observed heat flux-speed anti-correlation may result from an inverse relation between the escaping suprathermal fraction of the distribution and the size of the electric potential drop from the corona, which should scale with the solar wind speed \citep{scudder_theory_1979,scudder_theory_1979-1}. On the other hand, an increased fraction of suprathermals should itself lead to a higher wind speed in exospheric models \citep{maksimovic_kinetic_1997}, complicating the situation. Alternatively, the electron heat flux-speed anti-correlation we find near perihelion may simply reflect the initial conditions in the corona, similarly to the electron temperature-speed anti-correlation discussed above.   

\begin{figure}
\plotone{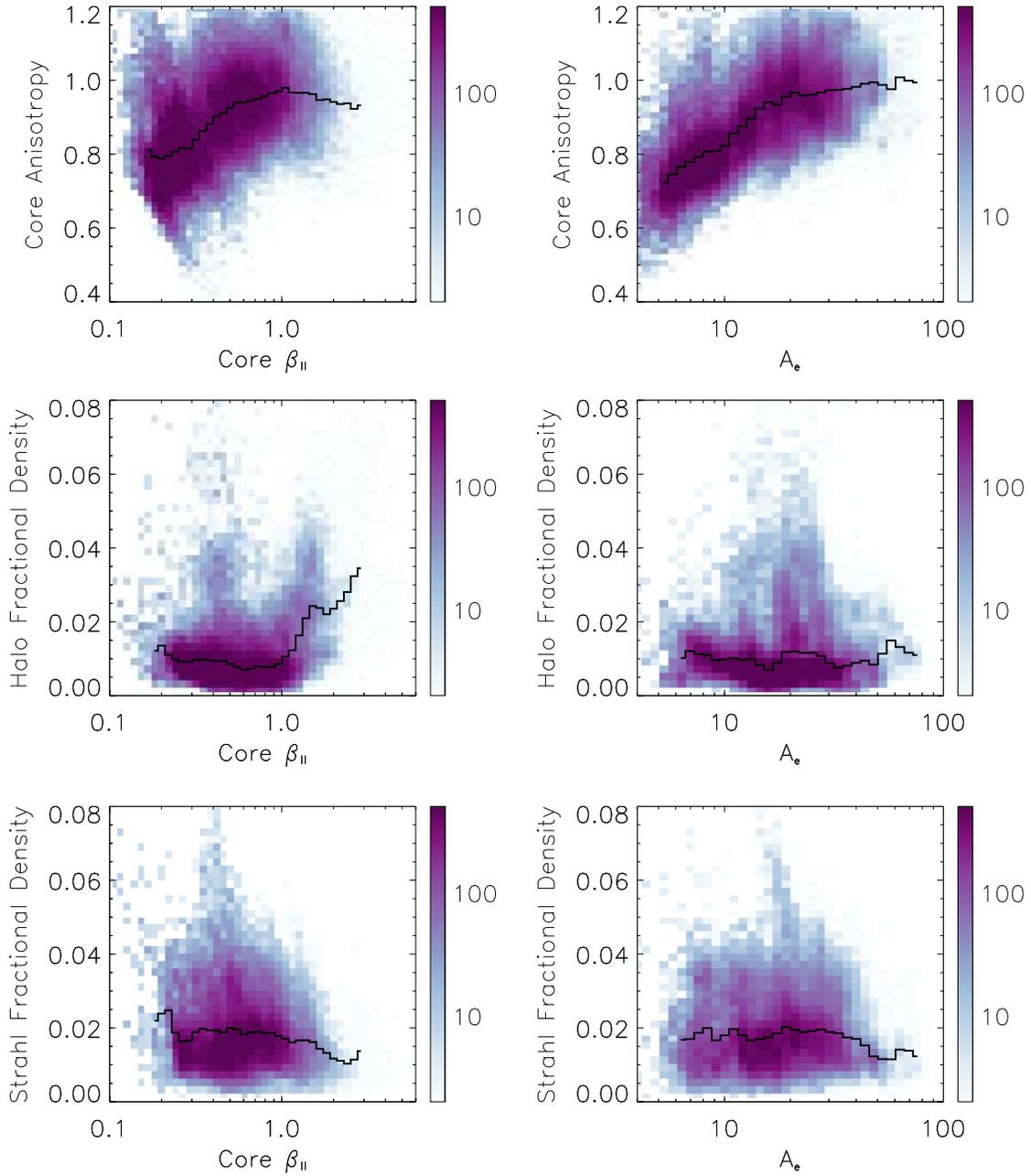}
\caption{Frequency distributions of selected distribution function parameters vs. $\beta_{c_{||}}$ (left column) and collisional age $A_e$ (right column) for all measurements within 0.25 AU, in the same format as Fig. \ref{fig:f4}. \label{fig:f5}}
\end{figure}

We next consider trends with respect to electron core $\beta_{||}$ and collisional age $A_e$, with selected results shown in Figure \ref{fig:f5}. We calculate collisional age in the same way as \cite{salem_electron_2003} and \cite{stverak_electron_2008}, but with the lower limit to the collision integral $r_0 = 0.1$ AU chosen to lie within the region of our measurements. For consistency with previous work, we utilize a temperature exponent $\alpha = 0.5$, also reasonably consistent with our temperature determinations. 

As expected, we find clear trends in core temperature anisotropy as a function of both $\beta_{||}$ and $A_e$, with more isotropic distributions observed for higher values of both parameters. These trends agree closely with previous results \citep{stverak_electron_2008}, and suggest that both instabilities and collisions act to isotropize the core even close to the Sun. 

On the other hand, though we tentatively observe some trends in both halo and strahl fractional density with $\beta_{||}$, we find at best weak trends in fractional density as a function of $A_e$. While in no way definitive given the very limited sampling of solar wind sources, these results may suggest that wave-particle interactions play a more important role than collisions in scattering the strahl electrons to form the halo close to the Sun. The recent model of \cite{vasko_whistler_2019} provides one possible instability mechanism that could limit the strahl fractional density at high $\beta_{||}$. 

\section{Conclusions and Implications} \label{sec:conc}

The first two orbits of PSP reveal a near-Sun electron environment in many ways similar to that observed at greater heliocentric distances. The electron distributions have the same basic form, with the same core, halo, and strahl populations typically observed at 1 AU. Furthermore, the observed radial scaling of the basic parameters of the distributions and the abundances of the core, halo, and strahl largely agrees with predictions from previous observations. As expected, near perihelion the strahl becomes narrower and dominates the suprathermal fraction of the distribution, and the halo almost disappears. Furthermore, the microphysics in operation near the Sun appears similar, with $\beta$ and $A_e$ controlling key parameters of the distribution in a manner similar to that observed at greater heliocentric distances. 

However, the near-Sun environment also holds surprises. The coronal electron temperature inferred from minor ion ratios anti-correlates with the asymptotic solar wind speed, but this correlation apparently disappears somewhere between the Sun and 1 AU. Our observations reveal that, near PSP's perihelion, an anti-correlation between electron temperature and wind speed remains apparent, potentially revealing a more pristine solar wind. Thanks to the still evident proton temperature-speed correlation, this also leads to a strong anti-correlation between the electron-proton temperature ratio and the wind speed near the Sun. Furthermore, near perihelion, we find that lower speed streams have electron distributions that carry greater heat flux, in contrast to observations at greater heliocentric distances. 

The electron temperature-speed anti-correlation suggests that the imprints of the initial conditions at the source of the solar wind remain apparent in the electron distributions near PSP's perihelion. This in turn implies that signatures of the initial solar wind acceleration mechanism may also remain evident in the PSP observations. If correct, our initial observations may present a challenge for purely exospheric models of solar wind acceleration, since in these models both higher electron temperature and higher fractions of suprathermal electrons at the corona should lead to a higher asymptotic wind speed. 

\acknowledgments
We acknowledge the SWEAP contract NNN06AA01C for support. 

\appendix
\section{Estimating Electron Core, Halo, and Strahl Parameters} \label{sec:appen}

The combined SPAN-A-E and SPAN-B-E FOV covers most of the sky, but has some gaps. Furthermore, in this work, we exclude the two highest deflection angle bins from either sensor, the two SPAN-A-E anodes with views closest to the rear of the spacecraft, and the two SPAN-B-E anodes with views closest to the deck, due to partial FOV obstructions. We therefore utilize a fitting procedure to characterize the observed core and halo electron distributions.

Before fitting model distributions to the observations, we utilize a flat-fielding procedure to correct for relative sensitivity variations over the sensor FOV, which can result from hemispheric non-concentricity, detector gain variations, and/or other factors. Previous investigations have demonstrated that the relative drifts of the core and suprathermal populations lie along the magnetic field to high accuracy \citep{pulupa_spin-modulated_2014}, so electron gyrotropy should hold in the solar wind proton frame, in the absence of discontinuities or small-scale gradients. Therefore, to determine the sensitivity as a function of anode for each sensor individually, and relative to each other, we impose electron gyrotropy around the magnetic field measured by FIELDS, in the proton frame determined by SPC. We repeat this analysis for each range of electron pitch angle and for different time ranges to find a single consistent set of time-independent relative sensitivities. The relative sensitivity so derived varies by over a factor of $\sim 5$ over the combined FOV, demonstrating the critical importance of this correction. 

We do not yet incorporate corrections for spacecraft magnetic and electric fields, though these may have significant effects on portions of the distributions \citep{Mcginnis:Correcting}.

To characterize the measured electron distributions, we utilize non-linear least squares fitting to a model distribution, employing a gradient expansion algorithm \citep{bevington_data_2002}. In practice, rather than fitting in units of distribution function $f$, we perform the fit in units of differential energy flux (proportional to $E^2f$), which scales with count rate. For each measurement, we perform multiple steps to retrieve the parameters of the distribution. We first obtain a core temperature estimate $kT_0$ by locating the energy of the peak differential energy flux and dividing by two (an exact calculation of the temperature for a purely Maxwellian distribution). We then fit to a model distribution consisting of an isotropic non-convecting Maxwellian $f_{sec}$ with a fixed temperature of $kT_{sec}=3.5$ eV representing the secondary electron contamination (Eq. \ref{equation:sec}), superposed with an anisotropic convecting bi-Maxwellian function $f_c$ representing the core (Eq. \ref{equation:core}). We allow the core distribution to drift along the magnetic field, but not perpendicular to it, with respect to the proton frame. We fit this two-component function to the observations over an energy range of 6 eV to $4kT_0$, for all pitch angles greater than $45^{\circ}$ from the strahl direction (determined by comparing the values of the distribution function parallel and anti-parallel to the magnetic field, at 314 eV), to obtain estimates of the core parameters. We also obtain estimates of the secondary electron density, which varies as expected as with both the core density and temperature.  

\begin{equation}\label{equation:sec}
    f_{sec}(v) = n_{sec} \left(\frac{m}{2 \pi k T_{sec}}\right)^{3/2}\exp{\left( \frac{-mv^2}{2kT_{sec}} \right)}
\end{equation}

\begin{equation}\label{equation:core}
    f_c(v_{\perp},v_{||}) = n_c \left(\frac{m}{2 \pi k}\right)^{3/2}\frac{1}{T_{\perp}\sqrt{T_{||}}}\exp{\left( \frac{-mv_{\perp}^2}{2kT_{\perp}} \right)}\exp{\left( \frac{-m(v_{||}-v_c)^2}{2kT_{||}} \right)}
\end{equation}

Next, we fit an isotropic non-convecting Maxwellian $f_h$ (Eq. \ref{equation:halo}) to the residual of the measurements with respect to the best-fit function derived above, over an energy range of $4kT_c$ to 2000 eV (where the total core temperature $T_c = (2T_{\perp}+T_{||})/3$), over the same non-strahl angular range, to obtain estimates of the halo parameters. We do not utilize energy steps for which the average count rate falls below the one-count level, which effectively removes any influence from instrumental backgrounds, since the average background rate lies well below the one-count level. 

\begin{equation}\label{equation:halo}
    f_{h}(v) = n_{h} \left(\frac{m}{2 \pi k T_{h}}\right)^{3/2}\exp{\left( \frac{-mv^2}{2kT_{h}} \right)}
\end{equation}

We found no need to consider a non-Maxwellian (e.g. a kappa/Lorentzian) distribution for either the core or halo, consistent with the results of \cite{maksimovic_radial_2005} and \cite{stverak_radial_2009}, who found near-Maxwellian halo distributions close to the Sun. 

Finally, we compute the density, field-aligned velocity, energy, and field-aligned heat flux moments of the strahl using Eqs. \ref{equation:nm}, \ref{equation:vm}, \ref{equation:em}, and \ref{equation:qm}, numerically integrating all positive residuals of the measurements $f_{res+}$ with respect to the superposition of the best-fit functions derived above, for energies above $4kT_c$ and pitch angles within $45^{\circ}$ of the strahl direction. Unlike the fit-based values for the core and halo parameters, the strahl parameters are subject to underestimates due to the presence of FOV gaps. 

\begin{equation}\label{equation:nm}
    n_s = \int{f_{res+}(v)d^3v}
    \end{equation}
\begin{equation}\label{equation:vm}
    <v_s> = \frac{1}{n_s}\int{v_{||}f_{res+}(v)d^3v}
\end{equation}
\begin{equation}\label{equation:em}
    <E_s> = \frac{1}{n_s}\int{\frac{1}{2}mv^2 f_{res+}(v)d^3v}
\end{equation}
\begin{equation}\label{equation:qm}
    Q_{s} = \int{\frac{1}{2}mv^2v_{||} f_{res+}(v)d^3v}
\end{equation}

To account for the anti-strahl deficit discussed in Section \ref{sec:data}, we compute partial density and velocity increments/decrements in the anti-strahl direction, numerically integrating the signed residuals $f_{res}$ of the measurements with respect to the model distributions using Eqs. \ref{equation:nm} and \ref{equation:vm}, but with $f_{res+}$ replaced by $f_{res}$, for energies above $4kT_c$ and pitch angles in the hemisphere opposite the strahl. The velocity moment in this case does not represent an average velocity in the typical sense, since the residual density moment can include both positive and negative contributions. 

\bibliography{references,prep}

\begin{thebibliography}{}
\expandafter\ifx\csname natexlab\endcsname\relax\def\natexlab#1{#1}\fi
\providecommand{\url}[1]{\href{#1}{#1}}

\bibitem[{{Bale} {et~al.}(2019)}]{Bale:prep}
{Bale}, S., {et~al.} 2019, accepted

\bibitem[{Bale {et~al.}(2016)Bale, Goetz, Harvey, Turin, Bonnell,
  Dudok de Wit, Ergun, MacDowall, Pulupa, Andre, Bolton, Bougeret, Bowen,
  Burgess, Cattell, Chandran, Chaston, Chen, Choi, Connerney, Cranmer,
  Diaz-Aguado, Donakowski, Drake, Farrell, Fergeau, Fermin, Fischer, Fox,
  Glaser, Goldstein, Gordon, Hanson, Harris, Hayes, Hinze, Hollweg, Horbury,
  Howard, Hoxie, Jannet, Karlsson, Kasper, Kellogg, Kien, Klimchuk,
  Krasnoselskikh, Krucker, Lynch, Maksimovic, Malaspina, Marker, Martin,
  Martinez-Oliveros, McCauley, McComas, McDonald, Meyer-Vernet, Moncuquet,
  Monson, Mozer, Murphy, Odom, Oliverson, Olson, Parker, Pankow, Phan,
  Quataert, Quinn, Ruplin, Salem, Seitz, Sheppard, Siy, Stevens, Summers,
  Szabo, Timofeeva, Vaivads, Velli, Yehle, Werthimer, \&
  Wygant}]{bale_fields_2016}
Bale, S.~D., Goetz, K., Harvey, P.~R., {et~al.} 2016, Space Science Reviews,
  204, 49

\bibitem[{Belcher \& Davis(1971)}]{belcher_large-amplitude_1971}
Belcher, J.~W., \& Davis, Leverett, J. 1971, Journal of Geophysical Research,
  76, 3534

\bibitem[{Berčič {et~al.}(2019)Berčič, Maksimović, Landi, \&
  Matteini}]{bercic_scattering_2019}
Berčič, L., Maksimović, M., Landi, S., \& Matteini, L. 2019, Monthly Notices
  of the Royal Astronomical Society, 486, 3404

\bibitem[{Bevington \& Robinson(2002)}]{bevington_data_2002}
Bevington, P., \& Robinson, D.~K. 2002, Data {Reduction} and {Error} {Analysis}
  for the {Physical} {Sciences}, 3rd edn. (Boston: McGraw-Hill Education)

\bibitem[{{Case} {et~al.}(2019)}]{Case:SPC}
{Case}, A.~C., {et~al.} 2019, accepted

\bibitem[{Feldman {et~al.}(1978)Feldman, Asbridge, Bame, Gosling, \&
  Lemons}]{feldman_characteristic_1978}
Feldman, W.~C., Asbridge, J.~R., Bame, S.~J., Gosling, J.~T., \& Lemons, D.~S.
  1978, Journal of Geophysical Research: Space Physics, 83, 5285

\bibitem[{Feldman {et~al.}(1975)Feldman, Asbridge, Bame, Montgomery, \&
  Gary}]{feldman_solar_1975}
Feldman, W.~C., Asbridge, J.~R., Bame, S.~J., Montgomery, M.~D., \& Gary, S.~P.
  1975, Journal of Geophysical Research, 80, 4181

\bibitem[{Fox {et~al.}(2016)Fox, Velli, Bale, Decker, Driesman, Howard, Kasper,
  Kinnison, Kusterer, Lario, Lockwood, McComas, Raouafi, \&
  Szabo}]{fox_solar_2016}
Fox, N.~J., Velli, M.~C., Bale, S.~D., {et~al.} 2016, Space Science Reviews,
  204, 7

\bibitem[{Gary {et~al.}(1975)Gary, Feldman, Forslund, \&
  Montgomery}]{gary_heat_1975}
Gary, S.~P., Feldman, W.~C., Forslund, D.~W., \& Montgomery, M.~D. 1975,
  Journal of Geophysical Research, 80, 4197

\bibitem[{Gary {et~al.}(1999)Gary, Skoug, \& Daughton}]{gary_electron_1999}
Gary, S.~P., Skoug, R.~M., \& Daughton, W. 1999, Physics of Plasmas, 6, 2607

\bibitem[{Geiss {et~al.}(1995)Geiss, Gloeckler, Steiger, Balsiger, Fisk,
  Galvin, Ipavich, Livi, McKenzie, Ogilvie, \& Et}]{geiss_southern_1995}
Geiss, J., Gloeckler, G., Steiger, R.~v., {et~al.} 1995, Science, 268, 1033

\bibitem[{Gloeckler {et~al.}(2003)Gloeckler, Zurbuchen, \&
  Geiss}]{gloeckler_implications_2003}
Gloeckler, G., Zurbuchen, T.~H., \& Geiss, J. 2003, Journal of Geophysical
  Research: Space Physics, 108, doi:10.1029/2002JA009286

\bibitem[{Graham {et~al.}(2017)Graham, Rae, Owen, Walsh, Arridge, Gilbert,
  Lewis, Jones, Forsyth, Coates, \& Waite}]{graham_evolution_2017}
Graham, G.~A., Rae, I.~J., Owen, C.~J., {et~al.} 2017, Journal of Geophysical
  Research: Space Physics, 122, 3858

\bibitem[{Hammond {et~al.}(1996)Hammond, Feldman, McComas, Phillips, \&
  Forsyth}]{hammond_variation_1996}
Hammond, C.~M., Feldman, W.~C., McComas, D.~J., Phillips, J.~L., \& Forsyth,
  R.~J. 1996, Astronomy and Astrophysics, 316, 350

\bibitem[{Hundhausen {et~al.}(1970)Hundhausen, Bame, Asbridge, \&
  Sydoriak}]{hundhausen_solar_1970}
Hundhausen, A.~J., Bame, S.~J., Asbridge, J.~R., \& Sydoriak, S.~J. 1970,
  Journal of Geophysical Research, 75, 4643

\bibitem[{Kasper {et~al.}(2016)Kasper, Abiad, Austin, Balat-Pichelin, Bale,
  Belcher, Berg, Bergner, Berthomier, \& Bookbinder}]{kasper_solar_2016}
Kasper, J.~C., Abiad, R., Austin, G., {et~al.} 2016, Space Science Reviews,
  204, 131

\bibitem[{{Kasper} {et~al.}(2019)}]{Kasper2019:prep}
{Kasper}, J.~C., {et~al.} 2019, accepted

\bibitem[{{Klein} {et~al.}(2019){Klein}, Martinovic, Huang, Halekas, Stevens,
  Alterman, Bale, Bowen, Case, Kasper, Korreck, Larson, Livi, MacDowell,
  Mallet, Paulson, Pulupa, \& Whittlesey}]{Klein:prep}
{Klein}, K.~G., Martinovic, M., Huang, J., {et~al.} 2019, in~prep.

\bibitem[{Lemaire \& Scherer(1971)}]{lemaire_kinetic_1971}
Lemaire, J., \& Scherer, M. 1971, Journal of Geophysical Research, 76, 7479

\bibitem[{Lemaire \& Scherer(1973)}]{lemaire_kinetic_1973}
---. 1973, Reviews of Geophysics, 11, 427

\bibitem[{Maksimovic {et~al.}(1997)Maksimovic, Pierrard, \&
  Lemaire}]{maksimovic_kinetic_1997}
Maksimovic, M., Pierrard, V., \& Lemaire, J.~F. 1997, Astronomy and
  Astrophysics, 324, 725

\bibitem[{Maksimovic {et~al.}(2005)Maksimovic, Zouganelis, Chaufray, Issautier,
  Scime, Littleton, Marsch, McComas, Salem, Lin, \&
  Elliott}]{maksimovic_radial_2005}
Maksimovic, M., Zouganelis, I., Chaufray, J.-Y., {et~al.} 2005, Journal of
  Geophysical Research: Space Physics, 110, doi:10.1029/2005JA011119

\bibitem[{Maxsimovic {et~al.}(2019)}]{Maxsimovic:Helios_comp}
Maxsimovic, M., {et~al.} 2019, in~prep.

\bibitem[{McComas {et~al.}(1992)McComas, Bame, Feldman, Gosling, \&
  Phillips}]{mccomas_solar_1992}
McComas, D.~J., Bame, S.~J., Feldman, W.~C., Gosling, J.~T., \& Phillips, J.~L.
  1992, Geophysical Research Letters, 19, 1291

\bibitem[{McFadden {et~al.}(2009)McFadden, Carlson, Larson, Bonnell, Mozer,
  Angelopoulos, Glassmeier, \& Auster}]{mcfadden_themis_2009}
McFadden, J.~P., Carlson, C.~W., Larson, D., {et~al.} 2009, in The {THEMIS}
  {Mission}, ed. J.~L. Burch \& V.~Angelopoulos (New York, NY: Springer New
  York), 477--508

\bibitem[{Mcginnis {et~al.}(2019)Mcginnis, Halekas, Whittlesey, Larson, \&
  Kasper}]{Mcginnis:Correcting}
Mcginnis, D., Halekas, J.~S., Whittlesey, P., Larson, D., \& Kasper, J.~C.
  2019, Journal of Geophysical Research: Space Physics

\bibitem[{Moncuquet {et~al.}(2019)}]{Moncuquet:QTN}
Moncuquet, M., {et~al.} 2019, in~prep.

\bibitem[{Pannekoek(1922)}]{pannekoek_ionization_1922}
Pannekoek, A. 1922, Bulletin of the Astronomical Institutes of the Netherlands,
  1, 107

\bibitem[{Pierrard \& Lemaire(1996)}]{pierrard_lorentzian_1996}
Pierrard, V., \& Lemaire, J. 1996, Journal of Geophysical Research: Space
  Physics, 101, 7923

\bibitem[{Pierrard {et~al.}(1999)Pierrard, Maksimovic, \&
  Lemaire}]{pierrard_electron_1999}
Pierrard, V., Maksimovic, M., \& Lemaire, J. 1999, Journal of Geophysical
  Research: Space Physics, 104, 17021

\bibitem[{Pilipp {et~al.}(1987)Pilipp, Miggenrieder, Montgomery, Mühlhäuser,
  Rosenbauer, \& Schwenn}]{pilipp_characteristics_1987}
Pilipp, W.~G., Miggenrieder, H., Montgomery, M.~D., {et~al.} 1987, Journal of
  Geophysical Research: Space Physics, 92, 1075

\bibitem[{Pilipp {et~al.}(1990)Pilipp, Miggenrieder, Mühläuser, Rosenbauer,
  \& Schwenn}]{pilipp_large-scale_1990}
Pilipp, W.~G., Miggenrieder, H., Mühläuser, K.-H., Rosenbauer, H., \&
  Schwenn, R. 1990, Journal of Geophysical Research: Space Physics, 95, 6305

\bibitem[{Pulupa {et~al.}(2014)Pulupa, Bale, Salem, \&
  Horaites}]{pulupa_spin-modulated_2014}
Pulupa, M.~P., Bale, S.~D., Salem, C., \& Horaites, K. 2014, Journal of
  Geophysical Research: Space Physics, 119, 647

\bibitem[{Rosenbauer {et~al.}(1977)Rosenbauer, Schwenn, Marsch, Meyer,
  Miggenrieder, Montgomery, Muehlhaeuser, Pilipp, Voges, \&
  Zink}]{rosenbauer_survey_1977}
Rosenbauer, H., Schwenn, R., Marsch, E., {et~al.} 1977, Journal of Geophysics
  Zeitschrift Geophysik, 42, 561

\bibitem[{Salem {et~al.}(2003)Salem, Hubert, Lacombe, Bale, Mangeney, Larson,
  \& Lin}]{salem_electron_2003}
Salem, C., Hubert, D., Lacombe, C., {et~al.} 2003, The Astrophysical Journal,
  585, 1147

\bibitem[{Scime {et~al.}(1994{\natexlab{a}})Scime, Bame, Feldman, Gary,
  Phillips, \& Balogh}]{scime_regulation_1994}
Scime, E.~E., Bame, S.~J., Feldman, W.~C., {et~al.} 1994{\natexlab{a}}, Journal
  of Geophysical Research: Space Physics, 99, 23401

\bibitem[{Scime {et~al.}(1994{\natexlab{b}})Scime, Phillips, \&
  Bame}]{scime_effects_1994}
Scime, E.~E., Phillips, J.~L., \& Bame, S.~J. 1994{\natexlab{b}}, Journal of
  Geophysical Research: Space Physics, 99, 14769

\bibitem[{Scudder(1992)}]{scudder_causes_1992}
Scudder, J.~D. 1992, The Astrophysical Journal, 398, 299

\bibitem[{Scudder(2019)}]{Scudder:SERM}
---. 2019, submitted.

\bibitem[{Scudder \& Olbert(1979{\natexlab{a}})}]{scudder_theory_1979}
Scudder, J.~D., \& Olbert, S. 1979{\natexlab{a}}, Journal of Geophysical
  Research: Space Physics, 84, 2755

\bibitem[{Scudder \& Olbert(1979{\natexlab{b}})}]{scudder_theory_1979-1}
---. 1979{\natexlab{b}}, Journal of Geophysical Research: Space Physics, 84,
  6603

\bibitem[{Tong {et~al.}(2015)Tong, Bale, Chen, Salem, \&
  Verscharen}]{tong_effects_2015}
Tong, Y., Bale, S.~D., Chen, C. H.~K., Salem, C.~S., \& Verscharen, D. 2015,
  The Astrophysical Journal Letters, 804, L36

\bibitem[{Vasko {et~al.}(2019)Vasko, Krasnoselskikh, Tong, Bale, Bonnell, \&
  Mozer}]{vasko_whistler_2019}
Vasko, I.~Y., Krasnoselskikh, V., Tong, Y., {et~al.} 2019, The Astrophysical
  Journal, 871, L29

\bibitem[{Vocks {et~al.}(2005)Vocks, Salem, Lin, \& Mann}]{vocks_electron_2005}
Vocks, C., Salem, C., Lin, R.~P., \& Mann, G. 2005, The Astrophysical Journal,
  627, 540

\bibitem[{{Whittlesey} {et~al.}(2019)}]{Whittlesey:SPANe}
{Whittlesey}, P., {et~al.} 2019, in~prep.

\bibitem[{Štverák {et~al.}(2009)Štverák, Maksimovic, Trávníček, Marsch,
  Fazakerley, \& Scime}]{stverak_radial_2009}
Štverák, S., Maksimovic, M., Trávníček, P.~M., {et~al.} 2009, Journal of
  Geophysical Research: Space Physics, 114, doi:10.1029/2008JA013883

\bibitem[{Štverák {et~al.}(2008)Štverák, Trávníček, Maksimovic, Marsch,
  Fazakerley, \& Scime}]{stverak_electron_2008}
Štverák, S., Trávníček, P., Maksimovic, M., {et~al.} 2008, Journal of
  Geophysical Research: Space Physics, 113, doi:10.1029/2007JA012733

\bibitem[{Štverák {et~al.}(2015)Štverák, Trávníček, \&
  Hellinger}]{stverak_electron_2015}
Štverák, S., Trávníček, P.~M., \& Hellinger, P. 2015, Journal of
  Geophysical Research: Space Physics, 120, 8177

\end{thebibliography}

\end{document}